\documentclass[%
 reprint,superscriptaddress,
 amsmath,amssymb,
 aps,prb,floatfix,longbibliography
]{revtex4-2}
\usepackage{comment}
\usepackage{hyperref}
\usepackage{color}
\usepackage{graphicx}
\usepackage{dcolumn}
\usepackage{bm}
\usepackage{here}
\usepackage{braket}
\usepackage{multirow}

\newcommand{\Tr}{\mathrm{Tr}}
\newcommand{\diagram}[2]{\;\vcenter{\hbox{\includegraphics[width=24mm,page=#2]{./#1.pdf}}}\;}
\DeclareUnicodeCharacter{2212}{-}
\begin{document}


\title{
Tensor Network Renormalization Study on the Crossover in\\
Classical Heisenberg and \texorpdfstring{$\mathrm{RP^2}$}{RP2} Models in Two Dimensions 
}

\author{Atsushi Ueda}
\email{aueda@issp.u-tokyo.ac.jp}
\affiliation{%
 Institute for Solid State Physics, University of Tokyo, Kashiwa 277-8581, Japan
}%
\author{Masaki Oshikawa}
\affiliation{%
 Institute for Solid State Physics, University of Tokyo, Kashiwa 277-8581, Japan
}%
\affiliation{Kavli Institute for the Physics and Mathematics of the Universe (WPI),
The University of Tokyo, Kashiwa, Chiba 277-8583, Japan}
\affiliation{Trans-scale Quantum Science Institute, University of Tokyo, Bunkyo-ku, Tokyo 113-0033, Japan}

\date{\today}

\begin{abstract}
We study the classical two-dimensional $\mathrm{RP^2}$ and Heisenberg models, using the Tensor-Network Renormalization (TNR) method.
The determination of the phase diagram of these models has been challenging and controversial due to the very large correlation lengths at low temperatures.
The finite-size spectrum of the transfer matrix obtained by TNR is useful in identifying the conformal field theory describing a possible critical point.
Our results indicate that the ultraviolet fixed point
for the Heisenberg model and the ferromagnetic $\mathrm{RP^2}$ model in the zero temperature limit
corresponds to a conformal field theory with central charge $c=2$, in agreement with two independent would-be Nambu-Goldstone modes.
On the other hand, the ultraviolet fixed point in the zero temperature limit
for the antiferromagnetic Lebwohl-Lasher model, which is a variant of the $\mathrm{RP^2}$ model,
seems to have a larger central charge. This is consistent with $c=4$ expected from the effective SO(5) symmetry.
At $T >0$, the convergence of the spectrum is not good in both the Heisenberg and ferromagnetic $\mathrm{RP^2}$ models.
Moreover, there seems to be no appropriate candidate of conformal field theory matching the spectrum, which shows the effective central charge $c \sim 1.9$.
These suggest that both models have a single disordered phase at finite temperatures, although the ferromagnetic $\mathrm{RP^2}$ model exhibits
a strong crossover at the temperature where the dissociation of $\mathbb{Z}_2$ vortices has been reported.
\end{abstract}
\maketitle
\section{Introduction}
The Berezinskii-Kosterlitz-Thouless (BKT) transition in the classical XY model is a milestone for studying topological phase transitions~\cite{Nobel2016,Berezinsky:1972rfj,Kosterlitz_1973,Kosterlitz_1974}. The $\mathrm{U(1)}$ spin configuration allows the topological defects (vortex-antivortex) with integer charge, and its phase transition mechanism is currently well understood~\cite{Kosterlitz_1974}.
In contrast to the remarkable success in the XY model, the nature of the topological dynamics with the other types of topological defects remains elusive.
The Lebwohl-Lasher (LL)~\cite{PhysRevA.6.426} and the antiferromagnetic Heisenberg model on the triangular lattice~\cite{kawamura1984phase,doi:10.1143/JPSJ.79.023701,doi:10.1143/JPSJ.79.084706} host $\mathbb{Z}_2$ vortices thanks to the $\mathbb{Z}_2$-valued fundamental group of the target space.
In analogy to the BKT transition, it is tempting to expect a topological phase transition driven by the $\mathbb{Z}_2$ vortices
with a quasi-long range order in the low-temperature phase. 
In fact, such a phase transition has been advocated and also supported by several numerical studies~\cite{PhysRevB.46.662,chiccoli1988monte,mondal2003finite,PhysRevE.70.066125,shabnam2016existence,fukugita1982numerical}.

However, even after more than 30 years, the existence of the phase transition is still inconclusive.
From the standpoint of the continuum effective field theory, the LL model for example would be described by a non-linear sigma model with the $\mathrm{RP^2}$
as the target space, in $1+1$ dimensions.
$\mathrm{RP^2}$ is simply the two-dimensional sphere $S^2$ with the antipodal points identified.
Thus, as far as the local geometry of the target space is concerned, the $\mathrm{RP^2}$ nonlinear sigma model is the same as the more standard O(3) nonlinear sigma model
(with $S^2$ as the target space).
Since the lowest-order term in the Renormalization Group (RG) beta function is given by the curvature of the target space~\cite{Friedan1980},
both the $\mathrm{RP^2}$ and 
the O(3) nonlinear sigma models are asymptotic-free~\cite{POLYAKOV197582,POLYAKOV1983121} (an infinitesimal coupling will be eventually renormalized to a strong coupling at large enough lengthscale).
In the context of statistical mechanics, the asymptotic freedom of the nonlinear sigma model suggests that the corresponding lattice model is disordered for
any non-zero temperature.
In fact, this is perhaps the standard picture on the phase diagram of the classical Heisenberg model\cite{sachdev2011quantum,fradkin2013field,altland2010condensed,PhysRevLett.36.691} defined by
\begin{align}
    H=-J\sum_{\langle i,j \rangle}\vec{S}_i\cdot\vec{S}_j,
\end{align}
where $\langle i,j \rangle$ runs over the pairs of nearest-neighbor sites $i$ and $j$, and 
$\vec{S}_i$ denotes a three-component unit vector. 
This is one of the most fundamental models in statistical mechanics.
However, even for the classical Heisenberg model, there is a controversy on the phase diagram, since a phase transition at finite temperature was
suggested based on numerical simulations~\cite{schmoll2021classical,brown19932d,kapikranian2007quasi,PhysRevB.2.684,PhysRevB.22.4462,alles1997perturbation,PhysRevD.59.067703} and high temperature expansions ~\cite{PhysRevLett.17.913,stanley2}.
While the nonlinear sigma model description looks reasonable, its applicability to the lattice model is not rigorously established and thus
it is possible that the lattice model behaves differently from the nonlinear sigma model prediction.
The phase diagram of the $\mathrm{RP^2}$ model~\cite{PhysRevLett.71.3906,PhysRevD.102.034513,caracciolo1994analytic,PhysRevD.53.3445,PhysRevD.53.5918,Delfino_2020,PhysRevE.78.051706} is even more controversial than that of the Heisenberg model. 
While the Monte Carlo simulations are very powerful for those classical spin systems, it is difficult to distinguish
a crossover from a phase transition because of
the extremely large correlation length~\cite{PhysRevE.78.051706,PhysRevLett.121.217801,PhysRevD.53.3445,PhysRevD.53.5918,PhysRevE.90.032109,PhysRev.181.811,PhysRevB.2.684,PhysRevB.22.4462,alles1997perturbation,PhysRevD.59.067703}.
\par{}In this paper, we reinvestigate the classical Heisenberg and $\mathrm{RP^2}$ models on the square lattice using the tensor network renormalization (TNR) scheme~\cite{PhysRevLett.99.120601,PhysRevLett.115.180405,PhysRevB.95.045117,PhysRevLett.118.110504,PhysRevLett.118.250602,PhysRevB.97.045111}.
Contrary to the conventional Monte-Carlo methods, TNR offers direct access to the spectrum of the transfer matrix and central charge of the conformal field theory (CFT), and it proves to be a powerful tool to study BKT transitions~\cite{Jha2020,PhysRevE.89.013308,PhysRevB.104.165132}. 
First we consider the $T\rightarrow0$ fixed point, where spontaneous symmetry breaking is not prohibited by the Mermin-Wagner theorem~\cite{PhysRevLett.17.1133}. 

Taking advantage of the TNR, we directly calculate the UV central charge of the Heisenberg model in the $T\sim0$ limit using TNR, to find that the central charge is $c=2$, which is strong evidence of the asymptotic freedom. 
We also analyze the $T\rightarrow0$ limit of the $\mathrm{RP^2}$ model in a similar manner. Interestingly, our numerical result indicates that the ultraviolet $T\rightarrow0$ $\mathrm{RP^2}$ fixed point also has the central charge $c=2$. On the other hand, the transfer matrix spectra of these two models have different structures, suggesting that the UV limits are distinct. Moreover, the possibility of the phase transition at finite temperature is also discussed.
Our results support a crossover rather than a true phase transition.
However, the observed strong crossover corresponds to a sudden change of the RG flow of the $\mathrm{RP^2}$ model in the vicinity of the reported transition temperature,
suggesting the existence of a repulsive fixed point near the actual RG trajectory.

\par{}
This paper is organized as follows. In Sec.~\ref{sec:models}, we first explain our models and their implementations in the tensor network in details. Then, we provide the methodology on how to extract the universal information from the renormalized tensors in Sec.~\ref{sec:methodology}. As our main results, the numerically obtained data for the zero and finite temperature fixed points are displayed in Sec.~\ref{sec:t0fixed_point} and \ref{sec:finiteT}, respectively. The interpretations and discussion thereof are given in Sec.\ref{sec:discussion}, and we summarize and refer to our future issues in Sec.~\ref{sec:conclusion}.

\section{Two-dimensional \texorpdfstring{$\mathrm{RP^2}$}{RP2} models}
\label{sec:models}

As discussed in the Introduction, we are interested in classical spin models with a spin ``target space'' $M$ for which $\pi_1(M)=\mathbb{Z}_2$.
This allows $\mathbb{Z}_2$ vortices as topological defects that are argued to induce a phase transition.
There are actually many different models which can host the $\mathbb{Z}_2$ vortices.
The simplest choice of the target space is $M= \mathrm{RP^2}$. We refer to such a spin model in general as a ``$\mathrm{RP^2}$ model''.
There are a few variations within the $\mathrm{RP^2}$ model.
The first is to consider the square of the Heisenberg interaction, and it is called the Lebwohl-Lasher (LL) model~\cite{PhysRevA.6.426},
\begin{align}
    H=-J\sum_{\langle i,j\rangle}(\vec{S}_i\cdot \vec{S}_j)^2,\label{LLmodel_eq}
\end{align}
where $\vec{S}_i$ is a three-component classical spin with $|\vec{S}_i|=1$, as in the Heisenberg model.
One can immediately check that this model is invariant to the local head-tail flip $\vec{S}_i \rightarrow - \vec{S}_i$ simultaneously at all the sites $i$.
There is plenty of research on the $J>0$ model, but less is found for $J<0$ in the literature.
We refer to these cases, respectively, as ``ferromagnetic'' and ``antiferromagnetic'' LL models.
In the case of the Heisenberg model, the ferromagnetic and antiferromagnetic models on the square lattice are equivalent, being mapped to each other by a redefinition
of the spin variables $\vec{S}_i \to - \vec{S}_i$ on the odd sites.
In contrast, ferromagnetic and antiferromagnetic LL models are not equivalent to each other.
In the antiferromagnetic LL model, the orthogonal alignment of neighboring ``spins'' is favored;
because of the freedom in choosing the orthogonal direction, the dimension of the ground-state manifold is larger in the antiferromagnetic LL model.

Another realization of the $\mathrm{RP^2} $ model is by introducing the Ising spins $\sigma_{ij}=\pm 1$ on the bonds~\cite{PhysRevD.58.074510,Caterall1998MontecarloRG},
which act as a $\mathbb{Z}_2$ gauge field on the lattice. The Hamiltonian is given by
\begin{align}
    H=-J\sum_{\langle i,j\rangle}\sigma_{ij}\vec{S}_i\cdot \vec{S}_j ,
    \label{RP2gauge_eq}
\end{align}
which we call the $\mathrm{RP^2}$ gauge model.

To compute the physical quantities by a tensor network, we decompose the local Boltzmann weights by the spherical harmonic and Legendre functions~\cite{PhysRevD.88.056005,Laurens-formalism} as,
\begin{align}
    e^{(\beta\cos(\gamma_{ij}))^2}&=\sum_lf_l(\beta)\sum_{m=-l}^{l}\bar{Y}_{lm}(\theta_i,\phi_i)Y_{lm}(\theta_j,\phi_j)\\
    f_l(\beta)&=2\pi\int_{-1}^{+1}dxP_l(x)e^{\beta x^2}\label{parity_LL}
\end{align}
where $(\theta_i,\phi_i)$ is the position of the spin on $S^2$ and $\gamma_{ij}$ is the angle between two spins.
This allows to rewrite the partition function in terms of the transfer matrix on each site.
\begin{align}
    Z&={\rm Tr}\prod_i\int\frac{d\Omega}{4\pi}\prod e^{(\beta\cos(\gamma_{ij}))^2}\\
    &={\rm Tr}\prod_{i\in \mathcal{L}}\sum_{l_i,m_i}f_{l_i}(\beta)\prod_s F_{l_1m_1,l_2m_2}^{l_3m_3,l_4m_4}
\end{align}
where 
\begin{align}
&F_{l_1m_1,l_2m_2}^{l_3m_3,l_4m_4}=\nonumber\\
&\frac{1}{4\pi}\sum_{l,m}G(l_1,l_2,l,m_1,m_2,m)G(l_3,l_4,l,m_3,m_4,-m)\nonumber
\end{align}
with $G$ being the gaunt function~\cite{gaunt1930continuous}.
Finally, one can express the local Boltzmann weights with a four-leg tensor as follows~\cite{Laurens-formalism}.
\begin{equation*}
    \diagram{transfermatrix}{1} = \left(\prod_{i=1}^4f_{l_i}(\beta)\right)^{\frac{1}{2}} F_{l_1,m_1,l_2,m_2}^{l_3,m_3,l_4,m_4}
\end{equation*}
The index $m_i$ takes the values in the range of $-l_i<m_i<l_i$ just like the magnetic quantum numbers. In practice, we introduce the cutoff for $l_i$ as $-l_{max}<l_i<l_{max}$ with its bond dimensions $(l_{max}+1)^2$ because the convergence of transfer matrix takes place for a large $l_{max}$. Moreover, as the head-tail symmetry prohibits $l_i$ from being odd, the bond dimension decreases to $\frac{1}{2}(l_{max}+1)(l_{max}+2)$. In the same way, one can construct the tensor network representation of the $\mathrm{RP^2} $ gauge model by changing $f_i(\beta)$ to
\begin{align}
    f_l(\beta)&=4\pi\int_{-1}^{+1}dxP_l(x)\cosh{(\beta x)}\label{parity_gauge}.
\end{align}
As the angular momentum cutoff, we use $l_{max}=4$ for the Heisenberg model and $l_{max}=6$ for both $\mathrm{RP^2}$ models. We note that the cutoffs above are not sufficient to calculate the accurate observables such as the free-energy density. Nonetheless, the truncation preserves the symmetry of the models, and thus we suspect that the universal information at criticality such as the central charge should remain intact under this truncation.
\section{Methodology}
\label{sec:methodology}
Critical points of statistical mechanical models correspond to RG fixed points.
In two dimensions, those RG fixed points are generally described by CFT in 1+1 dimensions.
Even if the system is off-critical, it could be described by a perturbed CFT when the correlation length is large.
Therefore, it is important to identify the CFT describing a possible critical point/phase, or underlying the region with a large correlation length.
The central charge $c$ is the most fundamental quantity characterizing a CFT.
Intuitively, it gives the ``number of degrees of freedom,'' so that free massless boson field theory has a central charge $1$ for each independent component.
Under certain general assumptions, Zamolodchikov proved that the central charge can only decrease along the RG flow~\cite{zamolodchikov1986irreversibility}.

In general, a two-dimensional statistical mechanical model, such as the Heisenberg model or the $\mathrm{RP^2} $ model studied in this paper,
can be mapped to a quantum many-body system in one spatial dimension.
The transfer matrix for the former corresponds to (the exponential of) the Hamiltonian of the latter.
This leads to many useful observations.
In particular, the TNR approach we employ naturally produces the spectrum of the transfer matrix,
which corresponds to the energy spectrum of the quantum many-body system in one spatial dimension.
When the system is critical and described by a CFT, the energy spectrum is dictated by the CFT.
In particular, it contains information on the central charge and scaling dimensions of operators.
The finite-size energy spectra of one-dimensional quantum systems have been extensively used to study quantum phases.
While it had been difficult to apply a similar method to two-dimensional statistical mechanical models, recently
a combination of the TNR and the finite-size scaling of the spectrum was demonstrated successfully for the two-dimensional XY model~\cite{PhysRevB.104.165132}.
In this paper, we extend the application of the TNR-based finite size scaling to the classical Heisenberg model and the $\mathrm{RP^2}$ models.
Since the analysis of the finite-size spectrum has been largely developed for one-dimensional quantum systems, we will
refer to the (the logarithm of) the eigenvalue spectrum of the finite-size transfer matrix as ``energy levels,'' referring to the quantum counterparts.
The $n$-th lowest energy eigenvalue $E_n(L)$ ($n=0,1,2,\ldots$) is determined as
\begin{equation}
    E_n(L) = - \frac{1}{L} \ln{\lambda_n(L)},
    \label{eq:def_en}
\end{equation}
where $\lambda_n(L)$ is the $n$-th largest eigenvalue of the renormalized tensor representing a $L \times L$ block contracted in the $x$-direction.
The (effective) central charge can be extracted from the scaling of the ``ground-state energy'' at each scale~\cite{cardy1984conformal}
\begin{equation}
E_0(L) \sim \epsilon_0 L-\frac{\pi c}{6L} .
\label{eq:c_FSS}
\end{equation}
Although the central charge is defined only for CFTs, we may define the effective central charge in terms of the scaling behavior of the ``ground-state energy'' $E_0$
as above in a given range of the scale.
Moreover, at a RG fixed point described by a CFT,
the scaling dimensions $x_n$ of operators in the CFT can be extracted from the scaling of the finite-size
``excitation energies''~\cite{PhysRevB.104.165132,cardy1984conformal} as
\begin{align}
\frac{2 \pi x_n}{L} = E_n(L) - E_0(L) .
\label{eq:def_xn}
\end{align}
Away from the RG fixed point, $x_n$ extracted from numerical data using this relation does not give a scaling dimension of a CFT.
However, again we often consider such a quantity $x_n$ as an effective scaling dimension, as it is useful in visualizing the RG flow.
\par
\begin{figure*}[tb]
    \centering
    \includegraphics[width=178mm]{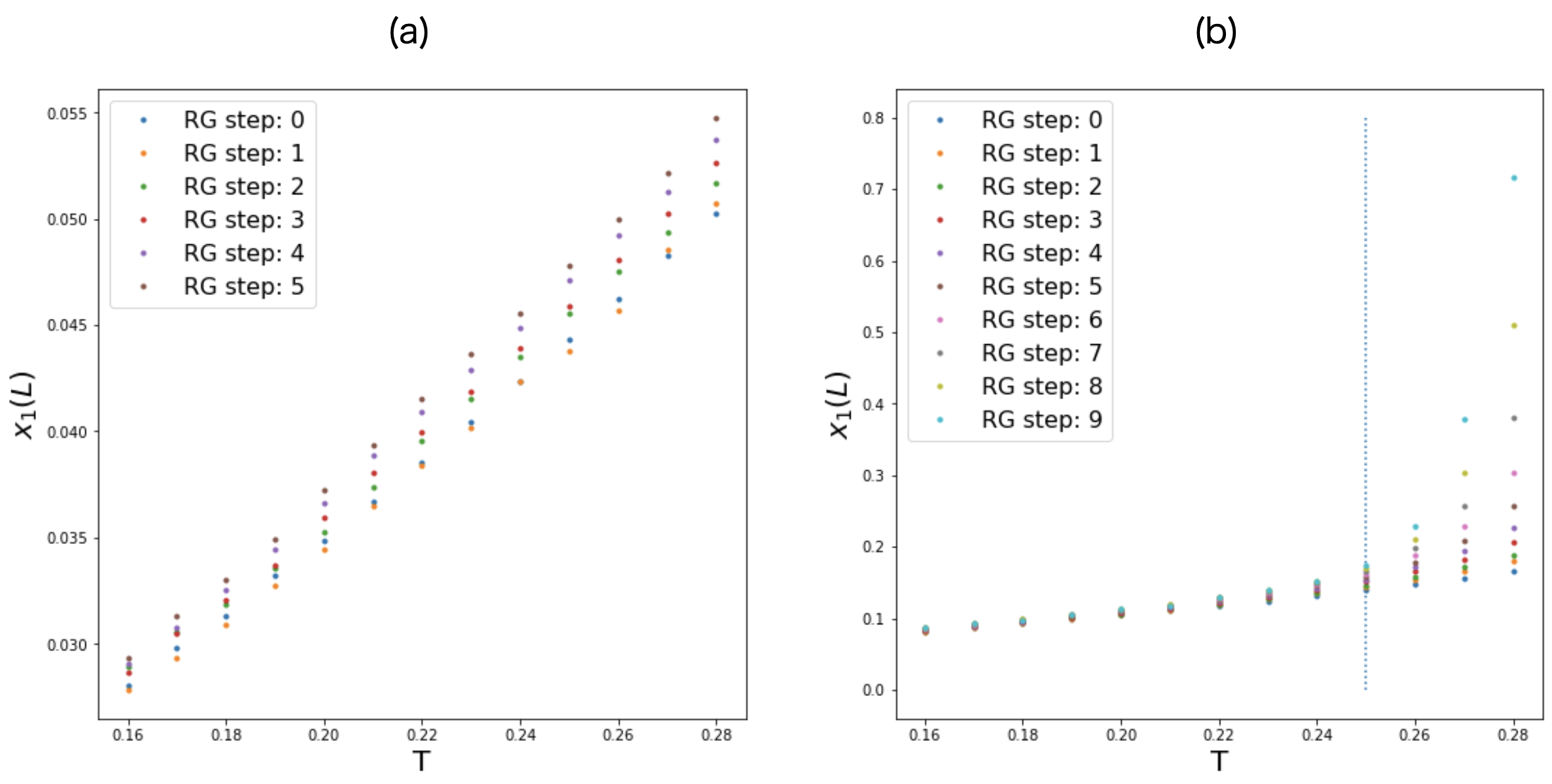}
    \caption{(a) The finite-size scaling dimension of the lowest excitation in the Heisenberg model. The scaling dimension grows monotonically along the RG steps, which is a consequence of the RG flow from the unstable $T=0$ fixed point. (b) The finite-size scaling dimension of the lowest excitation in the $\mathrm{RP^2} $ gauge model. The monotonic increase of $x_1(L)$ along the RG steps is observed in the whole range of temperature. The vertical dotted line denotes the reported transition temperature from previous studies, $T\sim0.25$.}
    \label{compare_flow}
\end{figure*}

{} In particular, $x_1(L)$ contains useful information about the correlation length $\xi(L)$. Using the eigenvalues of a $L\times L$ transfer matrix, the correlation length $\xi(L)$ can be represented as
\begin{align}
    \xi(L) = \frac{L}{\ln(\lambda_0(L)/\lambda_1(L))}.
\end{align}
Combining it with Eqs.~\eqref{eq:def_en} and \eqref{eq:def_xn}, we find the following relation between $x_1(L)$ and $\xi(L)$:
\begin{align}
    x_1(L) = \frac{1}{2\pi}\frac{L}{\xi(L)}.
\end{align}
At criticality, $\xi(L)$ diverges as $L$ increases, and $x_1(L)$ converges to a finite value, corresponding to the scaling dimension of its theory. On the other hand, $\xi(L)$ saturates for finitely-correlated systems, leading to the divergence of $x_1(L)$. Thus, it is a sign of a finite correlation length if $x_1(L)$ increases with RG steps.(Given the ground state is unique in the IR limit.) We carry out this analysis for finding the finite correlation length of the ${\rm RP^2}$ models shown in Figs.~\ref{compare_flow} and \ref{LL_fig}.
\section{Ultraviolet Fixed Point for \texorpdfstring{$T \rightarrow 0$}{T0}}
\label{sec:t0fixed_point}
The ground states of the Heisenberg or $\mathrm{RP^2} $ model are given by spins aligned in the same direction.
This means that we have a spontaneous symmetry breaking (SSB) at $T=0$.
On the other hand, since these models have continuous SO(3) symmetry,
the Mermin-Wagner theorem~\cite{PhysRevLett.17.1133} prohibits SSB at any finite temperature $T>0$.

It is interesting to consider the limit of $T \to +0$ from the finite temperature.
This has been studied for the O(3) nonlinear sigma model, which is believed to be the effective field theory for the classical Heisenberg model.
The SSB of continuous SO(3) symmetry in the $T \to 0$ limit implies the existence of the Nambu-Goldstone modes.
In the case of the O(3) nonlinear sigma model, the ``spin'' can fluctuate in two orthogonal directions around the ordered state.
Thus, the $T \to 0$ limit may be identified with the two-component free boson field theory with the central charge $c=2$.
The asymptotic freedom of the O(3) nonlinear sigma model implies that, at any $T>0$, the theory is described by the $c=2$ conformal field theory
with a relevant operator.
This means that the $c=2$ CFT is the ultraviolet (UV) fixed point, and the O(3) nonlinear sigma model corresponds to the RG flow emanating from it.
It is believed that the RG flow from the UV fixed point ends up in a massive theory, implying the absence of any intermediate phase or phase transitions in
the Heisenberg model at $T>0$.
This picture has been confirmed by the exact Bethe ansatz calculation of the free energy using the factorizable $S$-matrix
for the nonlinear sigma model~\cite{Fateev-Zam_ZN1991,Fendley-sigma2001} though the phase diagram for the Heisenberg model has yet to be identified~\cite{PhysRev.181.811,PhysRevB.2.684,PhysRevB.22.4462,alles1997perturbation,PhysRevD.59.067703}. 

\par{}On the other hand, the RG picture has not been clarified well for the $\mathrm{RP^2} $ model.
Since the target spaces of the $\mathrm{RP^2} $ and Heisenberg models are locally isomorphic to each other, we may also expect two-component $c=2$ free boson CFT
as the effective theory of the would-be Nambu-Goldstone modes in $T \to 0$.
However, to the best of our knowledge, this has not been studied for the $\mathrm{RP^2}$ model in Eqs.~\eqref{LLmodel_eq} and \eqref{RP2gauge_eq}.
Thus, here we present our TNR study on the ``effective central charge'' of the $\mathrm{RP^2} $ model.

\begin{figure*}[bt]
    \centering
    \includegraphics[width=178mm]{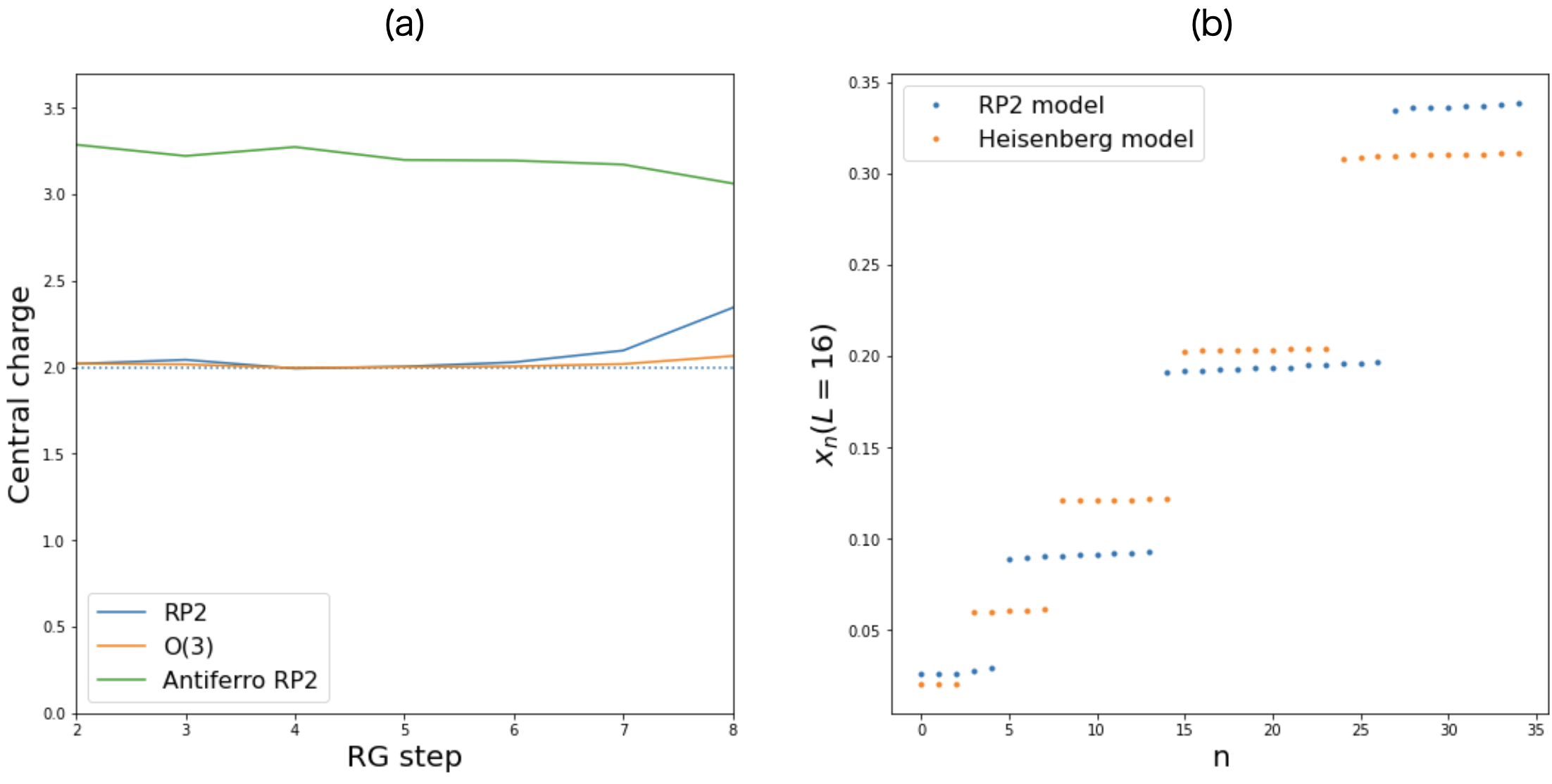}
    \caption{(a)Calculated central charge of the ferromagnetic LL, Heisenberg, and anti-ferromagnetic LL models at $T=0.005$. The ultra-violet central charge for the first two models is  $c=2$, whereas that of the AF-LL model is $c=4$. The numerically obtained data are $c=2.003$ and 1.998 for the Heisenberg and LL model, respectively, at the fourth RG step. The cusp around ten RG steps appears because we derive the central charge by a derivative of $L$ as $E_0(L)\sim \epsilon_0L-\frac{\pi c}{6L}$.[We only know $E_0(L)$ for $L=\sqrt{2}^n$, so linear fitting is needed.] The appearance of the cusps is a common feature when the system size reaches the correlation length. It could be the gap due to the finite bond dimensions. (b) The finite-size scaling dimension of the Heisenberg and LL model at $T=0.01$. The lowest excitation corresponds to spin 1 and 2, respectively. The degeneracy is slightly lifted due to the non-negligible finite-bond dimension effect at the large inverse temperature. To avoid unnecessary divergence of the Boltzmann weights, we shifted the energy in the computation, e.g., $H=-J\sum_{\langle i,j\rangle}[(\vec{S}_i\cdot \vec{S}_j)^2-1]$ for the $\mathrm{RP^2}$ LL model.}
    \label{UV_central}
\end{figure*}

Practically, it is challenging to simulate the limit $T \to 0$ due to the diverging Boltzmann weights.
Nevertheless, since the zero temperature fixed points should be RG unstable, observing the UV behavior(finite-size behavior) of $T\sim0$ allows us to access the information of the zero-temperature fixed point. 
Figure.~\ref{UV_central}(a) shows the central charge $c$ estimated with Eq.~\eqref{eq:c_FSS}
from Loop-TNR~\cite{PhysRevLett.118.110504} at $D=40$ and $T=0.005$.
We see that the effective central charge for both the Heisenberg model and the ferromagnetic LL model is stable at $c \sim 2$ at small RG steps.
As representative values, at $L=16$ (the fourth RG step),
we estimate $c=2.003$ and $c=1.998$, respectively, for the Heisenberg and the ferromagnetic LL model. 
This suggests that each of the systems behaves as a $c=2$ CFT at short distances.
In other words, the UV fixed point of these models is the $c=2$ CFT, as expected.
\begin{figure}[tb]
    \centering
    \includegraphics[width=86mm]{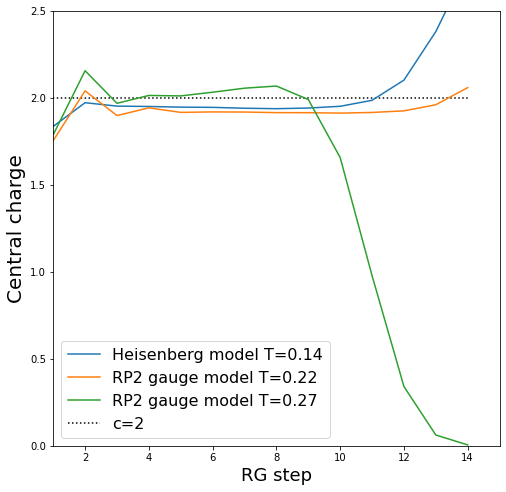}
    \caption{The effective central charge of the Heisenberg and $\mathrm{RP^2} $ gauge models at finite temperature. The numerically obtained central charge of the Heisenberg model is slightly smaller than 2, $c\sim 1.94$ at $T=0.14$ and the eighth RG step, implying that the correlation is large, but finite. Similarly, the central charge of the LL model is not converged and is $c\sim1.91$ at $T=0.22$ and the eighth RG step.  The cusps after the 12 RG steps are a numerical artifact due to the finite-bond dimension used in our calculation. On the other hand, at $T=0.27$ the effective central charge quickly drops to zero, which is a clear signature of the finite correlation length.}
    \label{compare_c}
\end{figure}

While the finite bond dimension effects are a limitation of the tensor-network type calculations in general,
the results are very accurate within lengthscales smaller than the bond-dimension induced correlation length.
The comparison with the CFT can be carried out for these lengthscales, as has been demonstrated in other contexts~\cite{PhysRevB.104.165132}. The detailed results of the finite bond dimension effects are summarized in Fig.~\ref{finiteD_plot}.
Therefore, we believe that our identification of the $c=2$ CFT as the UV fixed point for the low-temperature regime of the Heisenberg and ferromagnetic $\mathrm{RP^2} $ models
stands, in spite of the challenging nature of the problem. 
On the other hand, this observation does not directly answer the question about the nature of the ``transition'' to the disordered phase at higher temperature.

It is also interesting to study the antiferromagnetic LL model for comparison.
As mentioned in Sec.~\ref{sec:models}, the antiferromagnetic LL model has a higher-dimensional ground-state manifold.
It turns out that it corresponds to a O(5) ``spin'' model, as discussed in detail in Appendix~\ref{app:AF}. 
It is then expected that the UV fixed point is a $c=4$ free boson CFT, corresponding to four independent Nambu-Goldstone modes.
As shown in Fig.~\ref{UV_central}, we indeed find that the effective central charge of the antiferromagnetic LL model is significantly larger than the other models.
Although we do not observe the complete convergence, we think that the numerical result is consistent with our expectation of $c=4$ CFT representing the UV fixed point
representing the $T \to 0$ limit.

We also note that, Kawamura and Miyashita studied an antiferromagnetic Heisenberg model on the triangular lattice.
Effectively, this model would be described by a non-linear sigma model with the $\rm SO(3)$ group element as the field (principal chiral model).
Topologically, the $\rm SO(3)$ group is homeomorphic to $\mathrm{RP^3}$ and has the $\mathbb{Z}_2$ vortices.
We expect that the UV fixed point representing the $T \to 0$ limit of these models is a $c=3$ CFT.
It would be interesting to confirm this expectation numerically.

Meanwhile, it is noteworthy that the spectra of these two models are different in spite of the same central charge.
One can immediately see this in the effective scaling dimensions defined by Eq.~\eqref{eq:def_xn}.
In the field theoretical description, the scaling dimensions correspond to the energy levels of the collective excitation.
In particular, the lowest ones are three degenerate spin-1 excitations for the Heisenberg model.
In contrast, the $\rm O(3)$ spin is no longer the order parameter for the LL model under the local $\mathbb{Z}_2$ flip symmetry. Instead, we have nematic order parameter $Q$ defined as 
\begin{align}
    Q^{\alpha\beta}=s^\alpha s^\beta-\frac{1}{3}\delta^{\alpha\beta},
\end{align}
where $\alpha$ and $\beta$ runs through $(x,y,z)$. Thus, the lowest excitation of the LL model is spin 2 with its multiplicity 5. The numerically obtained finite-size scaling dimensions in Fig.~\ref{UV_central}(b) are consistent with the argument above.
This suggests that, although both are characterized by the same central charge $c=2$, the UV limits for the Heisenberg and $\mathrm{RP^2}$ models are distinct.

\section{\texorpdfstring{$\mathbb{Z}_2$}{Z2} vortex dissociation at \texorpdfstring{$T \sim T_*$}{TTstar}}
\label{sec:finiteT}

The main interest in the $\mathrm{RP^2} $ model is the possible topological phase transition due to the dissociation of the $\mathbb{Z}_2$ vortices, similarly to the
well-established BKT transition in the two-dimensional XY model~\cite{Nobel2016,Berezinsky:1972rfj,Kosterlitz_1973,Kosterlitz_1974}.
While there have been numerous studies supporting the existence of the transition, its nature has not been clarified.
Moreover, the very existence of the transition is still a question.

If the purported $\mathbb{Z}_2$ vortex dissociation transition is a second-order phase transition, it would be described by a CFT.
Likewise, if the system has a critical low-temperature phase similar to the low-temperature phase of the BKT transition, that phase would be also described by a CFT.
Hence, we have numerically estimated the effective central charge and scaling dimensions in the temperature region where the $\mathbb{Z}_2$ vortex dissociation is expected.

As a reference, we have also applied the same methodology to the classical Heisenberg model.
Figure.~\ref{compare_c} shows the effective central charge of the Heisenberg and $\mathrm{RP^2}$ gauge models at finite temperature, calculated from the TNR. 
First, the effective central charge of the Heisenberg model seems to have a plateau around $c \sim 2$ at $T=0.14$, which seems to suggest a critical phase.
However, the convergence of the effective central charge is not as good as in a similar TNR study of known critical points/phases
in the Ising and XY model~\cite{PhysRevB.104.165132,PhysRevLett.118.110504}. (Compare also with Fig.~\ref{UV_central}.)
In fact, a detailed analysis shows that the central charge is smaller than $c=2$ as is indicated with a black dotted line.
For example, the best estimate at $T=0.14$ for the Heisenberg model is $c = 1.94(1)$, obtained at the third through eighth RG steps.
Moreover, the ``energy levels'' which would give the scaling dimensions of the operators in the CFT change gradually, as shown in Fig.~\ref{compare_flow}.
As we will discuss in the next section, we believe that our results are more consistent with the standard ``asymptotic freedom'' picture, which
predicts the single disordered phase for $T>0$.

Now let us discuss the $\mathrm{RP^2}$ model around the purported $\mathbb{Z}_2$ vortex dissociation transition temperature $T_*=0.25$,
in a similar manner.
The effective central charge of the $\mathrm{RP^2}$ model also seems to have a plateau at $c \sim 2$ for $T = 0.22$ $(T < T_*)$, which seems to suggest a critical phase.
On the other hand, for $T = 0.27$ $(T > T_*)$, we find a quick drop of the effective central charge to zero, indicating a ``massive'' phase with a finite correlation length.
This implies a phase transition or a strong crossover around $T \sim T_* = 0.25$, in agreement with previous studies on the same model~\cite{PhysRevD.58.074510,Caterall1998MontecarloRG}.

\begin{figure}[tb]
    \centering
    \includegraphics[width=86mm]{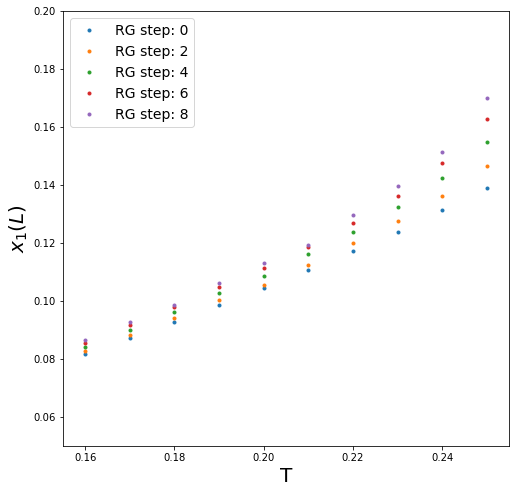}
    \caption{An enlargement of Fig.~\ref{compare_flow} (b) for the low-temperature regime $T \lesssim T_*$. The scaling dimension from the lowest excitation does not converge even here. }
    \label{fig:RP2_enlarged}
\end{figure}

On the other hand, we find that
the convergence of the effective central charge is again not good.
The best estimates of the central charge is $c=1.915(3)$ 
obtained at the fifth through eleventh RG steps for $T=0.22$ for the $\mathrm{RP^2}$ gauge model.
It is important to note that these values are smaller than $c=2$, which is expected for the UV fixed point, beyond numerical errors.
Moreover, the energy levels that would give the scaling dimensions of the operators in the CFT do not converge.
Below the ``transition temperature'' $T < T_*$, the change of the energy levels is certainly slower,
but it is still significant as shown in Fig.~\ref{fig:RP2_enlarged}.
In addition, the effective central charge and scaling dimensions of the LL model shown in Figs.~\ref{LL_fig} and ~\ref{fig:LL_enlarged} turn out to behave in a similar manner. Thus, we conclude that the both the $\rm RP^2$ gauge and LL models exhibit crossover rather than a true phase transition.

\section{Discussions}
\label{sec:discussion}
As noted in Sec.~\ref{sec:finiteT}, even though the effective central charge exhibits a ``plateau'' at $c \sim 1.9$
for both the Heisenberg and $\mathrm{RP^2}$ models at sufficiently low temperatures,
the convergence is not as good as in the similar TNR calculation for known critical points/phases.
If we still take the position that either of these models has a critical phase/point in those temperature regimes,
it would be described by a CFT with the central charge $c \sim 1.9 < 2$.
However, it is rather difficult to conceive such a CFT.

Considering that the model has a global SO(3) symmetry, it is tempting to consider the SO(3) Wess-Zumino-Witten (WZW) models.
SO(3) is locally isomorphic to SU(2), and the SU(2) WZW models are classified by the level $k$, which is a natural number.
The level-$k$ SU(2) WZW model has the central charge
\begin{equation}
    c = \frac{3k}{k+2} ,
\end{equation}
and the primary fields (with zero conformal spin) which have the scaling dimension
\begin{equation}
    x_j = 2 \frac{j (j+1)}{k+2} ,
\end{equation}
where $j$ is the SU(2) spin quantum number (half-odd-integer or integer), which satisfies $0 \leq j \leq k/2$.
An SO(3) WZW model corresponds to an SU(2) WZW model at an even level $k$.
The first nontrivial one is $k=2$ with $c=3/2$, which is too small for the observed value $c \sim 1.9$ in the Heisenberg and the $\mathrm{RP^2}$ models.
The next one is $k=4$ with $c=2$, which is too large.
One might be tempted to interpret the numerical estimate $c \sim 1.9$ (and poor convergence)
as a result of some numerical error, and postulate that the
system is actually described by the level-4 SU(2) model with $c=2$.
However, this does not seem to hold.

The ``energy levels'' of the lowest excited states [an SU(2) triplet]
found for the Heisenberg model do not converge up to the eighth RG step, as shown in Fig.~\ref{compare_flow}(a).
This also indicates that the system does not reach a RG fixed point yet.
Moreover, the observed energy level of the lowest excited state corresponds to the scaling dimension $x < 0.1$.
This is significantly smaller than the scaling dimension $x = 2/3$ of the corresponding triplet operator ($j=1$) in the level-4 SU(2) WZW model,
so that the identification is difficult.
Since our system does not contain a spinor representation of SU(2), we do not expect an SU(2) WZW model with an odd level $k$ to appears as
an effective model for the classical Heisenberg model.
Even if we disregard this issue and consider the level-3 SU(2) WZW model as a candidate CFT, it
also does not match the numerical observation both in terms of the central charge ($c=9/5=1.8$) and the spectrum (the lowest triplet operator $j=1$
having the scaling dimension $x=4/5$).
With these observations, we conclude instead that our numerical data support the standard ``asymptotic freedom'' scenario on the Heisenberg model
in which there is a single disordered phase for $T>0$.
The observed effective central charge $c \sim 1.9$ may reflect the RG flow from the $c=2$ UV fixed point.
This is consistent with the known exact RG flow~\cite{Fateev-Zam_ZN1991,Fendley-sigma2001}
of the O(3) nonlinear sigma model from $c=2$ to $c=0$ (``massive'' theory with a finite correlation length),
although its applicability to the Heisenberg model on the lattice could be questioned in principle.

We find a similar difficulty in interpreting the low-temperature regime $T \lesssim T_*$ of the $\mathrm{RP^2}$ model as a critical point/phase described by a CFT.
The lowest excited states in the $\mathrm{RP^2}$ form an SU(2) quintet, and their energy level corresponds to the scaling dimension $x \lesssim 0.2$.
This is again significantly smaller than the scaling dimension $x = 2$ of the corresponding quintet operator ($j=2$) in the level-4 SU(2) WZW model.
In fact, the convergence of the energy levels is much faster below $T \sim T_*$ in the $\mathrm{RP^2}$ model, compared to the high-temperature regime $T \gtrsim T_*$, as
shown in Fig.~\ref{LL_fig}(b). This indicates a remarkable change in the system around $T \sim T_*$, which is consistent with the $\mathbb{Z}_2$ vortex dissociation found
in other studies.
However, even in the low-temperature regime $T \lesssim T_*$, the energy levels do not show a true convergence as shown in Fig.~\ref{fig:LL_enlarged},
in contrast to the known critical point/phase.
Given the poor convergence of the effective central charge and the scaling dimensions, and also the lack of an appropriate CFT,
we conclude that it is more likely that there is no true phase transition at $T \sim T_*$ and that the region $T \lesssim T_*$ is ``massive,'' albeit with a large correlation length.
That is, similarly to what is widely believed for the Heisenberg model,
the $\mathrm{RP^2} $ model at finite temperature consists of the single disordered phase, which could be understood as an RG flow from the $c=2$ UV fixed point.
While we did observe a remarkable change in the behavior around $T \sim T_*$ where the dissociation of $\mathbb{Z}_2$ vortices has been reported, it would be understood as
a strong crossover rather than a true phase transition in this scenario.
\par{}Recently, the issue of the $\mathrm{RP^2}$ (and more generally $\mathrm{RP^{N-1}}$) model in two dimensions was studied in the framework of the scale-invariant scattering theory~\cite{2021Delfino}
The conclusion of those authors that there is no quasi-long-range ordered phase with varying exponents (RG fixed line) seems consistent with our findings.
On the other hand, their statement that the zero temperature fixed point for the $\mathrm{RP^{N-1}}$ model has an enlarged O[$M_N = N(N+1)/2-1$] symmetry may not be consistent with
our finding of $c=2$ on the $T \to 0$ UV fixed point for the $\mathrm{RP^2}$ model.
We note that $M_N=5$ for the $\mathrm{RP^2}$ model ($N=3$), and we indeed found an emergent O(5) symmetry for the ground states in the \emph{antiferromagnetic} LL model
(see Appendix~\ref{app:AF}), although we are not sure if our finding is related to the statement in Ref.~\cite{2021Delfino}.
More analysis is needed to clarify the relation between Ref.~\cite{2021Delfino} and our results.
In particular, it is desirable to identify the CFT corresponding to the ``additional solution'' for $N=3$ [Eq.~(32) of Ref.~\cite{2021Delfino}].
\section{Conclusions}
\label{sec:conclusion}
In this paper, we studied the low-temperature regimes of the Heisenberg and $\mathrm{RP^2}$ models using TNR, and we analyzed the spectrum of the finite-size transfer matrix.
When the system is critical and can be described by a CFT, the spectrum gives an estimate of the central charge and scaling dimensions of operators.
In the low-temperature limit $T \to 0$, we identify the UV fixed point of the Heisenberg and the ferromagnetic $\mathrm{RP^2}$ models as a $c=2$ CFT, consistently with the standard expectation.
On the other hand, the antiferromagnetic LL model indicates a larger value of the central charge of the UV fixed point. This is also consistent with the effective SO(5) symmetry of the
ground states, which predicts $c=4$ for the UV fixed point.

At moderately low temperatures of the Heisenberg model, we find the effective central charge $c \sim 1.9$ which looks somewhat stable over a range of the lengthscales up to $L=128$.
However, the finite-size energy levels, which give the scaling dimensions of operators in the underlying CFT, do not show convergence.
This and the lack of candidate CFTs supports the standard picture that the Heisenberg model has a single disordered phase with a finite correlation length for $T >0$.

The $\mathrm{RP^2}$ model in $T \lesssim T_*$ shows a similar behavior with the effective central charge $c \sim 1.9$.
However, the convergence is also not as good as in the spectrum of known critical points/phases, although it is faster than that in the $T \gtrsim T_*$ regime.
We conclude the it is more likely that the $\mathrm{RP^2}$ model also has a single disordered phase with a finite correlation length for $T>0$, based on the slow convergence and
the apparent lack of the candidate CFTs.

The phase diagrams of the Heisenberg and $\mathrm{RP^2}$ models have been controversial, chiefly due to the very large correlation length at low temperatures.
It is very difficult to distinguish directly a diverging correlation length from a large but finite one, in any numerical calculation.
On the other hand, analysis of the finite-size spectrum comparing with CFT has proved very powerful in quantum many-body systems in one spatial dimension;
the phase diagram can be determined precisely with an exact numerical diagonalization of the Hamiltonian of rather small systems.
This has been successfully extended to the two-dimensional classical XY model using the TNR~\cite{PhysRevB.104.165132}.
The present study demonstrates that, the finite-size spectrum from the TNR provides useful information also on the controversial problems of Heisenberg and $\mathrm{RP^2}$ models
despite the difficulties associated with the very large correlation lengths.

We do not completely rule out the possibility of the critical point at $T \sim T_*$ or the critical phase in $T \lesssim T_*$ suggested in several papers.
However, to pursue this viewpoint, one would need to explain the effective central charge and the spectra obtained in the present TNR study,
which is perhaps more difficult than simply discussing whether the correlation length diverges or not.
On the other hand, we do not have a complete explanation of what we believe is a strong crossover around $T \sim T_*$.
It is an interesting problem for the future to describe the crossover in terms of field theory,
for example with a CFT representing an RG fixed point in the vicinity and a relevant perturbation.

{\it\underline{Note added.}} Recently, we learned that Burgelman, Vanderstraeten, and Verstraete have also studied
the Heisenberg and LL model using the variational uniform MPS(VUMPS) method~\cite{burgelman2022contrasting}.
In that paper, they computed the high-accuracy correlation length from the transfer matrix and the effective central charge from the entanglement scaling. Detecting the deviation of the correlation length from the BKT scaling, they also concluded that this model does not go through a true phase transition. This is consistent with our diverging $x_1(L)$, which is a sign of a finite correlation length. Moreover, the effective central charge that they estimated is $c\sim 1.8$, which is consistent with our results of $c\sim 1.9$. The slight difference originates from the difference in the method for estimating the central charge, where we used the scaling of the ground-state energy. Nevertheless, this small discrepancy can be interpreted as additional evidence of a crossover because both methods should yield the same central charge at true criticality.
Concerning the nature of the fixed point that causes the crossover, they mention the similarity of the entanglement spectrum with the quantum
bilinear-biquadratic spin-1 Heisenberg chain. As future work, it would be interesting to compare it with the scaling dimensions found in our study.

\section*{Acknowledgements}

We thank Lander Burgelman, Laurens Vanderstraeten, and Frank Verstraete for sharing their draft prior to publication, and
for useful discussion on related subjects.
A. U. is supported by the MERIT-WINGS Program at the University of Tokyo, and by a JSPS Fellowship.
A part of the computation in this work has been done
using the facilities of the Supercomputer Center, the Institute for Solid State Physics, the University of Tokyo.
This work was supported in part by
MEXT/JSPS KAKENHI Grants No. JP17H06462, No. JP19H01808, and No. JP21J20523, and JST CREST Grant No. JPMJCR19T2

\appendix
\begin{figure*}[bt]
    \centering
    \includegraphics[width=178mm]{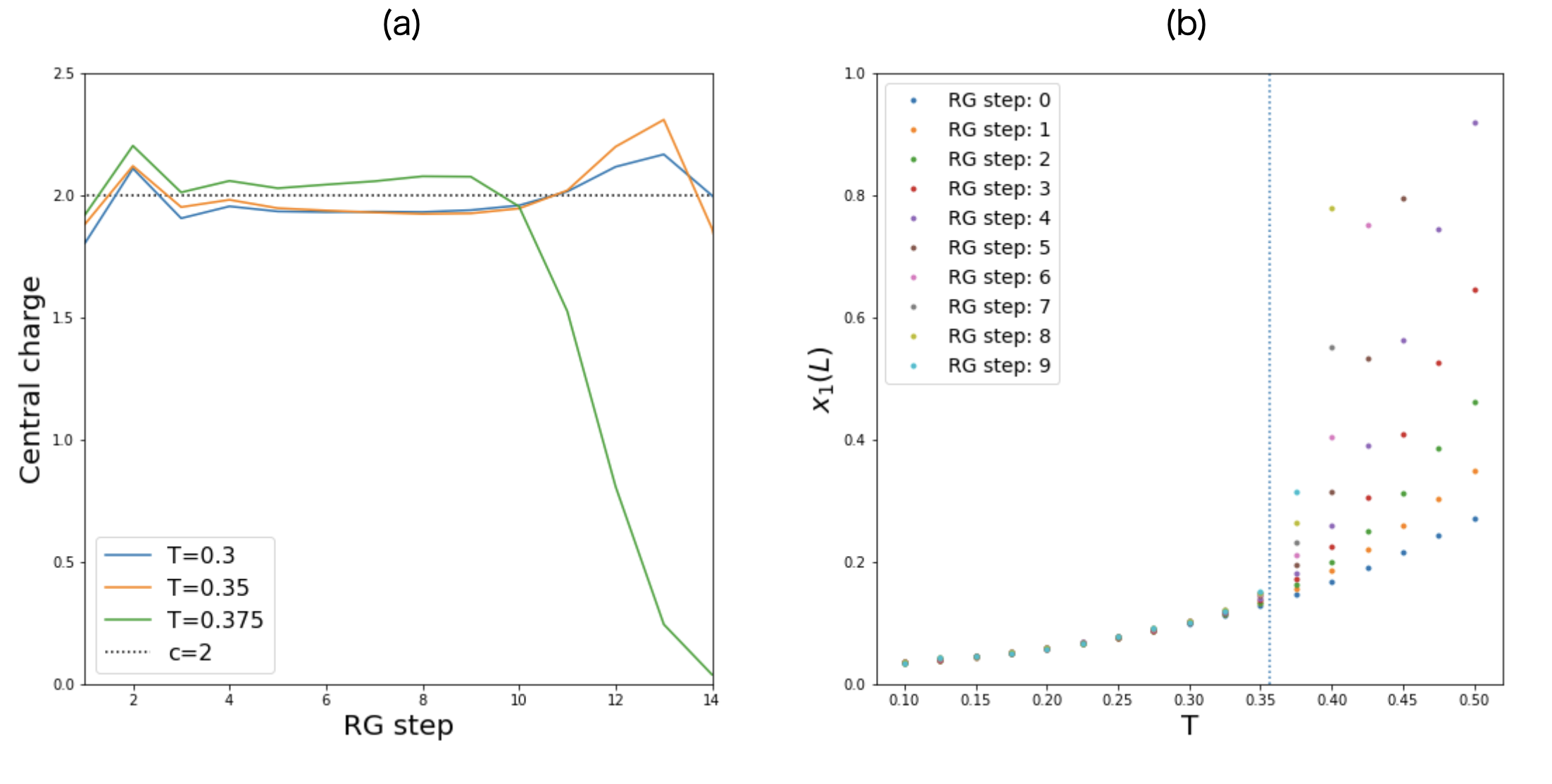}
    \caption{(a) The effective central charge of the Lebwohl-Lasher model at $T=0.3(T<T_*)$, $0.35(T\sim T_*)$, and $0.375(T>T_*)$. The numerically
obtained central charge of the Heisenberg model is slightly
smaller than 2, $c \sim 1.942(1)$ at $T = 0.35$ between the seventh and ninth RG steps, implying that the correlation is large, but finite. (b) The finite-size scaling dimension of the lowest excitation in the LL model. The monotonic increase of $x_1(L)$ along the RG steps is observed in the whole range of temperature. In particular, the RG flow becomes stronger around $T=0.35$. This phenomenon might be attributed to a repulsive fixed point in the vicinity of the trajectory of this model. The vertical dotted line denotes the transition temperature from previous studies $T=0.356$. }
    \label{LL_fig}
\end{figure*}
\begin{figure}[tb]
\begin{center}
\includegraphics[width=60mm]{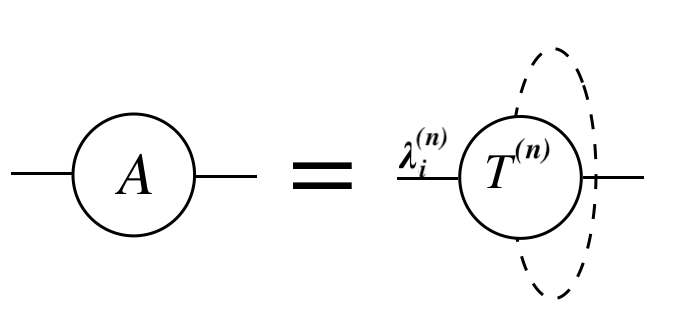}
\caption{A schematic picture of the procedure to extract the CFT information. It is possible to see the spectrum by making the $L\times L$ transfer matrix a cylinder.}
\label{acopy}
\end{center}
\end{figure}
\begin{figure}[tb]
    \centering
    \includegraphics[width=86mm]{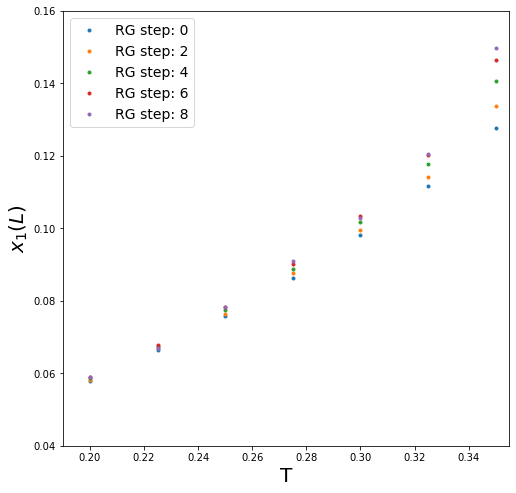}
    \caption{An enlargement of Fig.~\ref{LL_fig} (b) for the low-temperature regime $T \lesssim T_*$. The scaling dimension from the lowest excitation does not converge even here. }
    \label{fig:LL_enlarged}
\end{figure}
\begin{figure*}
    \centering
    \includegraphics[width=178mm]{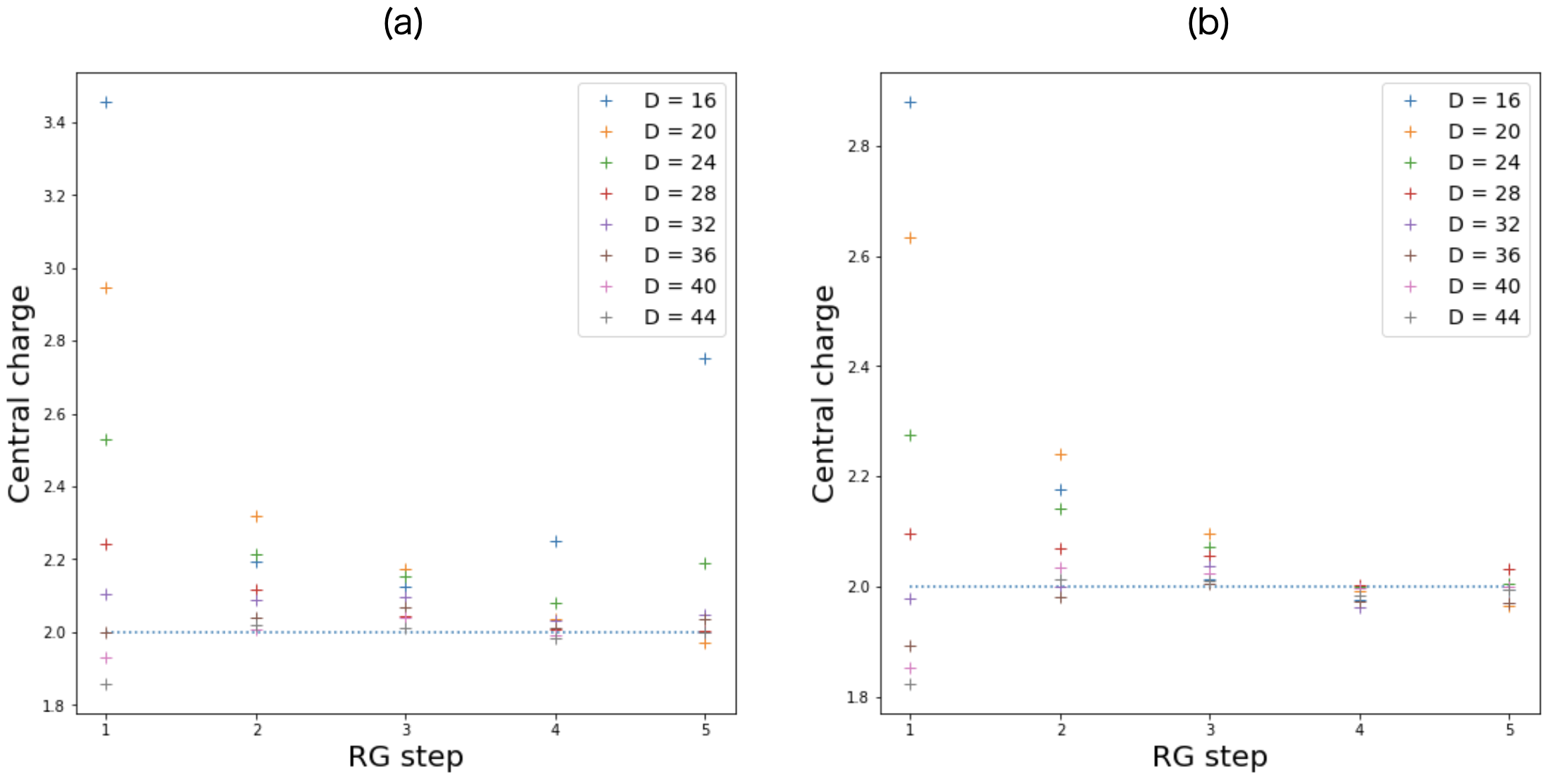}
    \caption{The ultraviolet central charge of (a) the Heisenberg and (b) LL model at $T=0.02$. The enormous finite bond dimension effect at $D=16$ decreases and the values converges at $D=44$.\label{finiteD_plot} }
\end{figure*}
\section{Effective central charge}
Ground state energy and excitation energies are written in the language of CFT as
\begin{align}
E_n-E_0&=\frac{2\pi}{L}x_n\label{a},\\
E_0&=\epsilon_0L-\frac{\pi c}{6L}\label{b},
\end{align}
where $x_n$ and $c$ are the scaling dimension and central charge, respectively. With real-space renormalization, the effective size of the renormalized tensor becomes $b$ times larger at each step.
Thus, starting from $L=1$, the effective scale ends up being $L=b^n$ after $n$ RG steps, and we denote the partition function at this scale as $Z(n)$. This partition function can be expressed with the generator along the $y$-axis using Eqs.~\eqref{a} and \eqref{b}.
\begin{align}
Z(n)&={\rm Tr} \exp(-b^nH_p)\nonumber\\
&={\rm Tr}_{x_i} \exp\left[-b^n(\frac{2\pi}{L}x_i-\frac{\pi c}{6L}+\epsilon_0L)\right]\nonumber\\
&={\rm Tr}_{x_i} \exp(-2\pi(x_i-\frac{c}{12})-\epsilon_0b^{2n})
\end{align}
This spectrum can be computed from the eigenvalues of the $L\times L$ transfer matrix constructed from the renormalized tensor $T^{(n)}$ as depicted in Fig.~\ref{acopy}. 
Assuming $x_0=0$, which corresponds to the identity operator, one is ready to recognize the relation between $\lambda_n$ and $x_n$.
\begin{align}
\frac{\lambda_n}{\lambda_0}=\exp(-2\pi x_n)
\end{align}
There is a pathological problem determining the central charge because we lack one equation. In order to eliminate the contribution from the bulk energy, or to normalize in other words, one needs to employ the partition function from previous step $Z(n-1)$.
\begin{align}
Z(n-1)={\rm Tr}_{x_i} \exp(-2\pi(x_i-\frac{c}{12})-\epsilon_0b^{2n-2})
\end{align}
For the fixed point tensor, it is reasonable to assume $c$ and $\epsilon_0$ to be constant. Then we get the central charge as below.
\begin{align}
c=\frac{6}{\pi}\frac{1}{b^2-1}(b^2\ln\lambda_0^{(n-1)}-\ln\lambda_0^{(n)})\label{c}
\end{align}
The formula widely used \cite{PhysRevB.80.155131}is 
\begin{align}
c=\frac{6}{\pi}\left[\frac{b^2}{b^2-1}\left(\ln Z(n-1)-\frac{\ln Z(n)}{b^2}\right)+\ln\frac{\lambda_0^{(n)}}{Z(n)}\right]\label{d}
\end{align}
Eqs.~\eqref{c} and \eqref{d} are equivalent in the critical case because $\frac{\lambda_0^{(n)}}{Z(n)}=\frac{\lambda_0^{(n-1)}}{Z(n-1)}$, though Eq.~\eqref{d} seems to be unstable when the system size exceeds the correlation length. Throughout this paper, we calculated the effective central charge with Eq.~\eqref{d}.\\

\section{Emergent \texorpdfstring{$O(5)$}{O5} universality in the AF-\texorpdfstring{$\mathrm{RP^2} $}{RP2} model}
\label{app:AF}

The emergence of the $\mathrm{O(5)}$ universality class seems peculiar. However, this phenomena is observed in three dimensions where the SSB of continuous symmetry is not prohibited \cite{PhysRevLett.17.1133}. The key idea is that the order parameter is no longer the $\mathrm{O(3)}$ spin, but the nematic order parameter $Q$ as 
\begin{align}
    Q^{\alpha\beta}=s^\alpha s^\beta-\frac{1}{3}\delta^{\alpha\beta}.
\end{align}
 As $Q^{\alpha\beta}$ is a symmetric trace-less rank-3 tensor, the number of independent components is five. This five components behave like a five component vector in Landau-Ginzburg-Wilson (LGW) theory, leading to the emergence of the $\mathrm{O(5)}$ universality~\cite{fernandez2005numerical,PhysRevB.71.014420}.
\par{} The ground state itself is different in the first place between $J>0$ and $J<0$. For ferromagnetic cases, the GS is fully aligned state $s_i\cdot s_j=\pm 1$. Except for the massive degeneracy due to the local $\mathbb{Z}_2$ gauge, it resembles that of the Heisenberg model. On the other hand, the ground state of the AF-$\mathrm{RP^2} $ model is non-trivial $(s_i\cdot s_j)=0$. Due this break in the symmetry between even and odd sites (sub-lattice), the order parameter might become a staggered one,
\begin{align}
A_i=(-1)^{x_i+y_i}Q_i.
\end{align}
Given the even-odd symmetry of $A_i\rightarrow-A_i$, LGW action for the AF-$\mathrm{RP^2} $ model become
\begin{align}
    F(A)=\int dr [(\nabla \Tr A)^2+a(\Tr A^2)+\nonumber\\
    b(\Tr A^2)^2+c(\Tr A^4)].
\end{align}
Let A be
\begin{align}
    A=
\begin{pmatrix}
a & c & d \\
c & b & e \\
d & e & -a-b\\
\end{pmatrix}.
\end{align}
Then $\Tr A^2=2(\Phi\cdot\Phi)$ and $\Tr A^2=2(\Phi\cdot\Phi)^2$, where $\Phi=(a+b/2,\frac{\sqrt{3}}{2}b,c,d,e)$. This is simply the $\mathrm{O(5)}$ model. Unlike the three dimensional model, $\phi^6$ is also relevant in 2D. However, the most stable fixed point(smallest central charge) of $\Phi^{2n}$ theory is that of $\Phi^4$, so our argument might be correct. Note that this LGW approach is justifiable only at $T=0$, where SSB is allowed under continuous symmetry. The verification of $c=4$ with numerical simulation is almost impossible due to the enormous bond dimension effect. Nonetheless, the central charge is larger than 3, implying the shadow of $c=4$.
In fact, our conjecture is consistent with the recent study of the $\mathrm{RP^2}$ model based on the conformal scattering theory, and this $\mathrm{O(5)}$ universality class corresponds to the solution $\mathrm{A_{+}}$ of Ref.~\cite{2021Delfino}, while the remaining solution $\mathrm{A_{-}}$ corresponds to the $c=\frac{5}{2}$ $\mathrm{SO(5)_1}$ WZW~\cite{2021CFTscatter,francesco2012conformal}. Interestingly, the central charge $c=2$ of the ferromagnetic $\mathrm{RP^2}$ model at zero temperature is none of them above, leaving the possibility for it to become a candidate for the solution $\mathrm{B_{\pm3}}$.
\bibliography{manuscript}
\end{document}